\pdfoutput=1
\documentclass[12pt,a4paper]{article}

\usepackage{ifthen} 
\newboolean{pdflatex}
\setboolean{pdflatex}{true} 

\newboolean{articletitles}
\setboolean{articletitles}{true} 

\newboolean{uprightparticles}
\setboolean{uprightparticles}{false} 

\newboolean{inbibliography}
\setboolean{inbibliography}{false} 

\def\paperauthors{LHCb collaboration} 
\def\paperasciititle{Measurement of branching fraction ratios for B+->D*+D-K+, B+->D*-D+K+, and B0->D*-D0K+ decays} 
\def\papertitle{Measurement of branching fraction ratios for \BpDstpDmKp, \BpDstmDpKp, \\ and \BzDstmDzKp decays} 
\def\paperkeywords{{High Energy Physics}, {LHCb}} 
\def\papercopyright{\the\year\ CERN for the benefit of the LHCb collaboration} 
\def\paperlicence{CC BY 4.0 licence}
\def\paperlicenceurl{https://creativecommons.org/licenses/by/4.0/}

\usepackage[top=1in, bottom=1.25in, left=1in, right=1in]{geometry}

\usepackage{microtype}
\usepackage{lineno}  
\usepackage{xspace} 
\usepackage{caption} 

\usepackage{graphicx}  
\usepackage{color}
\usepackage{colortbl}
\graphicspath{{./figs/}} 

\usepackage{amsmath} 
\usepackage{amssymb}
\usepackage{amsfonts}
\usepackage{upgreek} 

\newcommand*\patchAmsMathEnvironmentForLineno[1]{%
\expandafter\let\csname old#1\expandafter\endcsname\csname #1\endcsname
\expandafter\let\csname oldend#1\expandafter\endcsname\csname
end#1\endcsname
 \renewenvironment{#1}%
   {\linenomath\csname old#1\endcsname}%
   {\csname oldend#1\endcsname\endlinenomath}%
}
\newcommand*\patchBothAmsMathEnvironmentsForLineno[1]{%
  \patchAmsMathEnvironmentForLineno{#1}%
  \patchAmsMathEnvironmentForLineno{#1*}%
}
\AtBeginDocument{%
\patchBothAmsMathEnvironmentsForLineno{equation}%
\patchBothAmsMathEnvironmentsForLineno{align}%
\patchBothAmsMathEnvironmentsForLineno{flalign}%
\patchBothAmsMathEnvironmentsForLineno{alignat}%
\patchBothAmsMathEnvironmentsForLineno{gather}%
\patchBothAmsMathEnvironmentsForLineno{multline}%
\patchBothAmsMathEnvironmentsForLineno{eqnarray}%
}

\usepackage{hyperxmp}

\usepackage[pdftex,
            pdfauthor={\paperauthors},
            pdftitle={\paperasciititle},
            pdfkeywords={\paperkeywords},
            pdfcopyright={Copyright (C) \papercopyright},
            pdflicenseurl={\paperlicenceurl}]{hyperref}

\usepackage[colorinlistoftodos,textsize=scriptsize]{todonotes}

\usepackage[bottom,flushmargin,hang,multiple]{footmisc}

\usepackage[all]{hypcap} 

\usepackage{xspace} 
\usepackage{upgreek}

\def\DstarmKtpi  {{\ensuremath{\D^{*-}_{K3\pi}}}\xspace}
\def\DstarpKtpi   {{\ensuremath{\D^{*+}_{K3\pi}}}\xspace}
\def\DstarmKpi  {{\ensuremath{\D^{*-}_{K\pi}}}\xspace}
\def\DstarpKpi   {{\ensuremath{\D^{*+}_{K\pi}}}\xspace}

\def\DzKtpi         {{\ensuremath{\D^0_{K3\pi}}}\xspace}
\def\DzKpi         {{\ensuremath{\D^0_{K\pi}}}\xspace}

\def\DzbKtpi       {{\ensuremath{\Dbar{}^0_{K3\pi}}}\xspace}
\def\DzbKpi       {{\ensuremath{\Dbar{}^0_{K\pi}}}\xspace}

\def\Ncorr {{\ensuremath{N^{\mathrm{corr}}}}\xspace}

\def\BpDstpDmKp           {\decay{\Bu}{\Dstarp\Dm\Kp}}
\def\BpDstpK3piDmKp       {\decay{\Bu}{\DstarpKtpi\Dm\Kp}}
\def\BpDstpKpiDmKp       {\decay{\Bu}{\DstarpKpi\Dm\Kp}}
\def\BpDstmDpKp           {\decay{\Bu}{\Dstarm\Dp\Kp}}
\def\BpDstmK3piDpKp           {\decay{\Bu}{\DstarmKtpi\Dp\Kp}}
\def\BpDstmKpiDpKp           {\decay{\Bu}{\DstarmKpi\Dp\Kp}}
\def\BzDstmDzKp           {\decay{\Bd}{\Dstarm\Dz\Kp}}
\def\BzDstmKpiDzKpiKp           {\decay{\Bd}{\DstarmKpi\DzKpi\Kp}}
\def\BzDstmK3piDzKp    {\decay{\Bd}{\DstarmKtpi\Dz\Kp}}
\def\BzDstmK3piDzKpiKp    {\decay{\Bd}{\DstarmKtpi\DzKpi\Kp}}
\def\BzDstmDzK3piKp    {\decay{\Bd}{\Dstarm\DzKtpi\Kp}}
\def\BzDstmKpiDzK3piKp    {\decay{\Bd}{\DstarmKpi\DzKtpi\Kp}}
\def\BzDmDzKp              {\decay{\Bd}{\Dm\Dz\Kp}}
\def\BzDmDzKpiKp              {\decay{\Bd}{\Dm\DzKpi\Kp}}
\def\BzDmDzK3piKp       {\decay{\Bd}{\Dm\DzKtpi\Kp}}
\def\BpDzbarK3piDzKpiKp   {\decay{\Bu}{\DzbKtpi\DzKpi\Kp}}
\def\BpDzbarKpiDzK3piKp   {\decay{\Bu}{\DzbKpi\DzKtpi\Kp}}
\def\BpDzbarDzKp          {\decay{\Bu}{\Dzb\Dz\Kp}}

\def\BpDstpKtpiDmKp       {\decay{\Bu}{\DstarpKtpi\Dm\Kp}}
\def\BpDstmKtpiDpKp           {\decay{\Bu}{\DstarmKtpi\Dp\Kp}}

\def\BzDstmKtpiDzKpiKp    {\decay{\Bd}{\DstarmKtpi\DzKpi\Kp}}

\def\BzDstmKpiDzKtpiKp    {\decay{\Bd}{\DstarmKpi\DzKtpi\Kp}}
\def\BzDmDzKtpiKp       {\decay{\Bd}{\Dm\DzKtpi\Kp}}
\def\BpDzbarKtpiDzKpiKp   {\decay{\Bu}{\DzbKtpi\DzKpi\Kp}}
\def\BpDzbarKpiDzKtpiKp   {\decay{\Bu}{\DzbKpi\DzKtpi\Kp}}

\def\BDstDK           {\decay{\B}{\Dstar\Dbar\kaon}}
\def\BDDK           {\decay{\B}{\D\Dbar\kaon}}

\def\lhcb {\mbox{LHCb}\xspace}

\def\babar  {\mbox{BaBar}\xspace}

\def\cleo   {\mbox{CLEO}\xspace}

\def\aleph  {\mbox{ALEPH}\xspace}

\def\lhc    {\mbox{LHC}\xspace}

\def\MagUp {\mbox{\em Mag\kern -0.05em Up}\xspace}

\ifthenelse{\boolean{uprightparticles}}%
{

 \def\Ppi         {\ensuremath{\uppi}\xspace}

 \def\PDelta      {\ensuremath{\Delta}\xspace}                 
 \def\PXi      {\ensuremath{\Xi}\xspace}                 
 \def\PLambda      {\ensuremath{\Lambda}\xspace}                 
 \def\PSigma      {\ensuremath{\Sigma}\xspace}                 
 \def\POmega      {\ensuremath{\Omega}\xspace}                 
 \def\PUpsilon      {\ensuremath{\Upsilon}\xspace}                 
 
 \def\PB      {\ensuremath{\mathrm{B}}\xspace}                 
                  
 \def\PD      {\ensuremath{\mathrm{D}}\xspace}

 \def\PK      {\ensuremath{\mathrm{K}}\xspace}

 \def\PW      {\ensuremath{\mathrm{W}}\xspace}

 \def\Pb      {\ensuremath{\mathrm{b}}\xspace}                 
 \def\Pc      {\ensuremath{\mathrm{c}}\xspace}

 \def\Pi      {\ensuremath{\mathrm{i}}\xspace}

 \def\Pp      {\ensuremath{\mathrm{p}}\xspace}

 \def\Ps      {\ensuremath{\mathrm{s}}\xspace}

}
{

 \def\Ppi         {\ensuremath{\pi}\xspace}

 \mathchardef\PDelta="7101
 \mathchardef\PXi="7104
 \mathchardef\PLambda="7103
 \mathchardef\PSigma="7106
 \mathchardef\POmega="710A
 \mathchardef\PUpsilon="7107
                  
 \def\PB      {\ensuremath{B}\xspace}                 
                  
 \def\PD      {\ensuremath{D}\xspace}

 \def\PK      {\ensuremath{K}\xspace}

 \def\PW      {\ensuremath{W}\xspace}

 \def\Pb      {\ensuremath{b}\xspace}                 
 \def\Pc      {\ensuremath{c}\xspace}

 \def\Pi      {\ensuremath{i}\xspace}

 \def\Pp      {\ensuremath{p}\xspace}

 \def\Ps      {\ensuremath{s}\xspace}

}

\makeatletter
\ifcase \@ptsize \relax
  \newcommand{\miniscule}{\@setfontsize\miniscule{4}{5}}
\or
  \newcommand{\miniscule}{\@setfontsize\miniscule{5}{6}}
\or
  \newcommand{\miniscule}{\@setfontsize\miniscule{5}{6}}
\fi
\makeatother

\DeclareRobustCommand{\optbar}[1]{\shortstack{{\miniscule (\rule[.5ex]{1.25em}{.18mm})}
  \\ [-.7ex] $#1$}}

\def\W      {{\ensuremath{\PW}}\xspace}

\def\squark    {{\ensuremath{\Ps}}\xspace}
\def\squarkbar {{\ensuremath{\overline \squark}}\xspace}

\def\cquark    {{\ensuremath{\Pc}}\xspace}

\def\bquark    {{\ensuremath{\Pb}}\xspace}
\def\bquarkbar {{\ensuremath{\overline \bquark}}\xspace}
\def\bbbar     {{\ensuremath{\bquark\bquarkbar}}\xspace}

\def\pion   {{\ensuremath{\Ppi}}\xspace}

\def\pip    {{\ensuremath{\pion^+}}\xspace}
\def\pim    {{\ensuremath{\pion^-}}\xspace}

\def\kaon    {{\ensuremath{\PK}}\xspace}
  \def\Kbar    {{\kern 0.2em\overline{\kern -0.2em \PK}{}}\xspace}

\def\KorKbar    {\kern 0.18em\optbar{\kern -0.18em K}{}\xspace}
\def\Kz      {{\ensuremath{\kaon^0}}\xspace}

\def\Kp      {{\ensuremath{\kaon^+}}\xspace}
\def\Km      {{\ensuremath{\kaon^-}}\xspace}

  \def\Dbar    {{\kern 0.2em\overline{\kern -0.2em \PD}{}}\xspace}
\def\D       {{\ensuremath{\PD}}\xspace}
\def\Db      {{\ensuremath{\Dbar}}\xspace}
\def\DorDbar    {\kern 0.18em\optbar{\kern -0.18em D}{}\xspace}
\def\Dz      {{\ensuremath{\D^0}}\xspace}
\def\Dzb     {{\ensuremath{\Dbar{}^0}}\xspace}
\def\Dp      {{\ensuremath{\D^+}}\xspace}
\def\Dm      {{\ensuremath{\D^-}}\xspace}

\def\Dstar   {{\ensuremath{\D^*}}\xspace}
\def\Dstarpar   {{\ensuremath{\D^{(*)}}}\xspace}

\def\Dstarparb  {{\ensuremath{\Dbar{}^{(*)}}}\xspace}
\def\Dstarz  {{\ensuremath{\D^{*0}}}\xspace}

\def\Dstarp  {{\ensuremath{\D^{*+}}}\xspace}
\def\Dstarm  {{\ensuremath{\D^{*-}}}\xspace}

\def\B       {{\ensuremath{\PB}}\xspace}
\def\Bbar    {{\ensuremath{\kern 0.18em\overline{\kern -0.18em \PB}{}}}\xspace}

\def\BorBbar    {\kern 0.18em\optbar{\kern -0.18em B}{}\xspace}
\def\Bz      {{\ensuremath{\B^0}}\xspace}

\def\Bu      {{\ensuremath{\B^+}}\xspace}

\def\Bp      {{\ensuremath{\Bu}}\xspace}

\def\Bd      {{\ensuremath{\B^0}}\xspace}

  \def\Y#1S{\ensuremath{\PUpsilon{(#1S)}}\xspace}

\def\proton      {{\ensuremath{\Pp}}\xspace}

\def\Lbar        {{\ensuremath{\kern 0.1em\overline{\kern -0.1em\PLambda}}}\xspace}
\def\LorLbar    {\kern 0.18em\optbar{\kern -0.18em \PLambda}{}\xspace}

\def\BF         {{\ensuremath{\mathcal{B}}}\xspace}

\newcommand{\decay}[2]{\ensuremath{#1\!\to #2}\xspace}         

\def\to                 {\ensuremath{\rightarrow}\xspace}

\def\AT#1     {\ensuremath{A_{\mathrm{T}}^{#1}}\xspace}           

\def\C#1      {\ensuremath{\mathcal{C}_{#1}}\xspace}                       
\def\Cp#1     {\ensuremath{\mathcal{C}_{#1}^{'}}\xspace}                    
\def\Ceff#1   {\ensuremath{\mathcal{C}_{#1}^{\mathrm{(eff)}}}\xspace}        
\def\Cpeff#1  {\ensuremath{\mathcal{C}_{#1}^{'\mathrm{(eff)}}}\xspace}       
\def\Ope#1    {\ensuremath{\mathcal{O}_{#1}}\xspace}                       
\def\Opep#1   {\ensuremath{\mathcal{O}_{#1}^{'}}\xspace}                    

\newcommand{\tev}{\ifthenelse{\boolean{inbibliography}}{\ensuremath{~T\kern -0.05em eV}}{\ensuremath{\mathrm{\,Te\kern -0.1em V}}}\xspace}
\newcommand{\gev}{\ensuremath{\mathrm{\,Ge\kern -0.1em V}}\xspace}
\newcommand{\mev}{\ensuremath{\mathrm{\,Me\kern -0.1em V}}\xspace}
\newcommand{\kev}{\ensuremath{\mathrm{\,ke\kern -0.1em V}}\xspace}
\newcommand{\ev}{\ensuremath{\mathrm{\,e\kern -0.1em V}}\xspace}
\newcommand{\gevc}{\ensuremath{{\mathrm{\,Ge\kern -0.1em V\!/}c}}\xspace}
\newcommand{\mevc}{\ensuremath{{\mathrm{\,Me\kern -0.1em V\!/}c}}\xspace}
\newcommand{\gevcc}{\ensuremath{{\mathrm{\,Ge\kern -0.1em V\!/}c^2}}\xspace}
\newcommand{\gevgevcccc}{\ensuremath{{\mathrm{\,Ge\kern -0.1em V^2\!/}c^4}}\xspace}
\newcommand{\mevcc}{\ensuremath{{\mathrm{\,Me\kern -0.1em V\!/}c^2}}\xspace}

\def\mm   {\ensuremath{\mathrm{ \,mm}}\xspace}

\def\mum  {\ensuremath{{\,\upmu\mathrm{m}}}\xspace}

\def\invfb   {\ensuremath{\mbox{\,fb}^{-1}}\xspace}

\def\ps   {\ensuremath{{\mathrm{ \,ps}}}\xspace}

\newcommand{\chisq}{\ensuremath{\chi^2}\xspace}

\newcommand{\chisqip}{\ensuremath{\chi^2_{\text{IP}}}\xspace}

\def\gsim{{~\raise.15em\hbox{$>$}\kern-.85em
          \lower.35em\hbox{$\sim$}~}\xspace}
\def\lsim{{~\raise.15em\hbox{$<$}\kern-.85em
          \lower.35em\hbox{$\sim$}~}\xspace}

\def\sPlot{\mbox{\em sPlot}\xspace}

\def\ptot       {\mbox{$p$}\xspace}
\def\pt         {\mbox{$p_{\mathrm{ T}}$}\xspace}

\def\evtgen     {\mbox{\textsc{EvtGen}}\xspace}

\def\geant      {\mbox{\textsc{Geant4}}\xspace}

\def\photos     {\mbox{\textsc{Photos}}\xspace}

\def\pythia     {\mbox{\textsc{Pythia}}\xspace}

\def\tell1  {TELL1\xspace}
\def\ukl1   {UKL1\xspace}

\newcommand{\ie}{\mbox{\itshape i.e.}\xspace}

\newcommand{\etc}{\mbox{\itshape etc.}\xspace}

\usepackage{cite} 
\usepackage{mciteplus}

\usepackage{longtable} 

\usepackage{mwe}
\usepackage{amsmath}
\usepackage{mathrsfs}
\usepackage{booktabs}
\usepackage{multirow}

\usepackage{comment}

\begin{document}

\renewcommand{\thefootnote}{\fnsymbol{footnote}}
\setcounter{footnote}{1}


\begin{titlepage}
\pagenumbering{roman}

\vspace*{-1.5cm}
\centerline{\large EUROPEAN ORGANIZATION FOR NUCLEAR RESEARCH (CERN)}
\vspace*{1.5cm}
\noindent
\begin{tabular*}{\linewidth}{lc@{\extracolsep{\fill}}r@{\extracolsep{0pt}}}
\ifthenelse{\boolean{pdflatex}}
{\vspace*{-1.5cm}\mbox{\!\!\!\includegraphics[width=.14\textwidth]{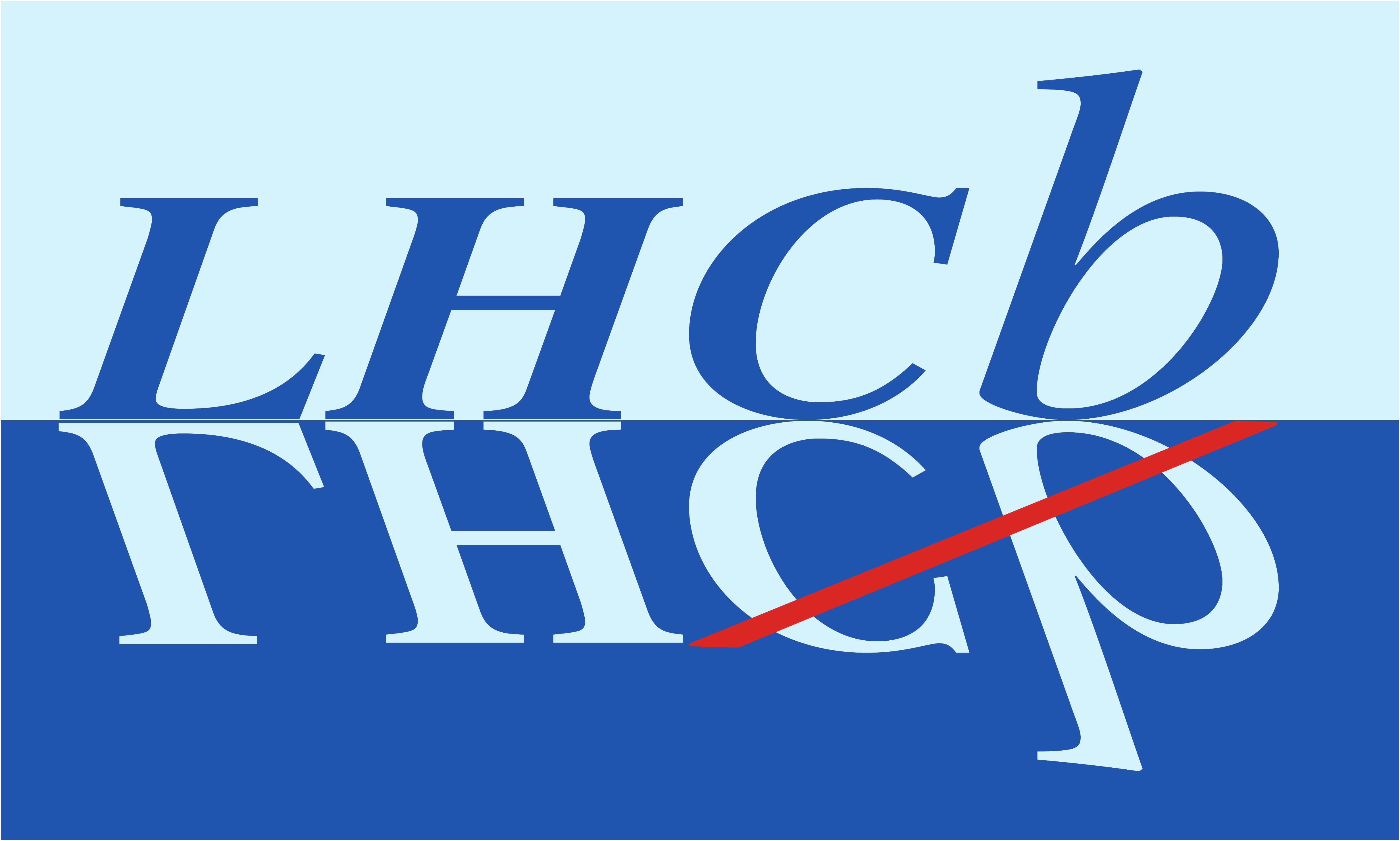}} & &}%
{\vspace*{-1.2cm}\mbox{\!\!\!\includegraphics[width=.12\textwidth]{figs/lhcb-logo.eps}} & &}%
\\
 & & CERN-EP-2020-056 \\  
 & & LHCb-PAPER-2020-006 \\  
 & & Jan 6, 2021 \\ 
 & & \\
\end{tabular*}

\vfill

{\normalfont\bfseries\boldmath\huge
\begin{center}
  \papertitle 
\end{center}
}

\vfill

\begin{center}
\paperauthors\footnote{Authors are listed at the end of this paper.}
\end{center}

\vfill

\begin{abstract}
  \noindent
 A measurement of four branching-fraction ratios for three-body decays of \B mesons involving two open-charm hadrons in the final state is presented. Run~1 and Run~2 \proton\proton collision data are used, recorded by the \lhcb experiment at centre-of-mass energies $7$, $8$, and $13\tev$ and corresponding to an integrated luminosity of $9\invfb$. The measured branching-fraction ratios are
\begin{equation*}
\begin{split}
\frac{\BF (\BpDstpDmKp)}{\BF (\BpDzbarDzKp)} &= 0.517 \pm 0.015 \pm 0.013 \pm 0.011  , \\[10pt]
\frac{\BF (\BpDstmDpKp)}{\BF (\BpDzbarDzKp)} &= 0.577 \pm 0.016 \pm 0.013 \pm 0.013  , \\[10pt]
\frac{\BF (\BzDstmDzKp)}{\BF (\BzDmDzKp)}    &= 1.754 \pm 0.028 \pm 0.016 \pm 0.035  , \\[10pt]
\frac{\BF (\BpDstpDmKp)}{\BF (\BpDstmDpKp)}  &= 0.907 \pm 0.033 \pm 0.014  ,
\end{split}
\end{equation*}
where the first of the uncertainties is statistical, the second systematic, and the third is due to the uncertainties on the \D-meson branching fractions. These are the most accurate measurements of these ratios to date. 
  
\end{abstract}

\vfill

\begin{center}
Published in JHEP 12 (2020) 139
\end{center}

\vfill

{\footnotesize 
\centerline{\copyright~\papercopyright. \href{\paperlicenceurl}{\paperlicence}.}}
\vspace*{2mm}

\end{titlepage}


\newpage
\setcounter{page}{2}
\mbox{~}
%
%
%
%


\renewcommand{\thefootnote}{\arabic{footnote}}
\setcounter{footnote}{0}

\cleardoublepage


\pagestyle{plain} 
\setcounter{page}{1}
\pagenumbering{arabic}


\section{Introduction}
\label{sec:Introduction}
 
There is a long history of studies of $\B \to \Dstarpar \Dstarparb K$ decays, where \B represents a \Bp or a \Bz meson, \Dstarpar is a \Dz, \Dstarz, \Dp, or \Dstarp meson, \Dstarparb is a charge conjugate of one of the \Dstarpar mesons, and \kaon is either a \Kp or \Kz meson.\footnote{The inclusion of charge conjugated processes  is implied throughout, unless otherwise stated.} The first observations of $\B \to \Dstarpar \Dstarparb K$ decays were made public in 1997 and 1998 by the \cleo~\cite{cleoconf} and \aleph~\cite{Barate:1998ch} collaborations. They fully reconstructed a number of these decay modes in order to probe the discrepancy between the measured values of branching fractions for hadronic and semileptonic decays of the \B meson~\cite{Bigi:1993fm}, the at that time unresolved `charm-counting problem'. In 2003, the \babar collaboration published the first comprehensive investigation of $\B \to \Dstarpar \Dstarparb K$ decays, reporting observations or limits for 22 channels~\cite{Aubert:2003jq}. Later, in 2011, the measurements were updated using a  five times larger data sample~\cite{delAmoSanchez:2010pg}. The \lhcb data collected during Run 1 and Run 2 of the Large Hadron Collider (\lhc) provide an opportunity to obtain an order of magnitude larger yields with smaller backgrounds than those measured previously.

\begin{figure}[tbp]
\centering
\includegraphics[width=0.885\linewidth]{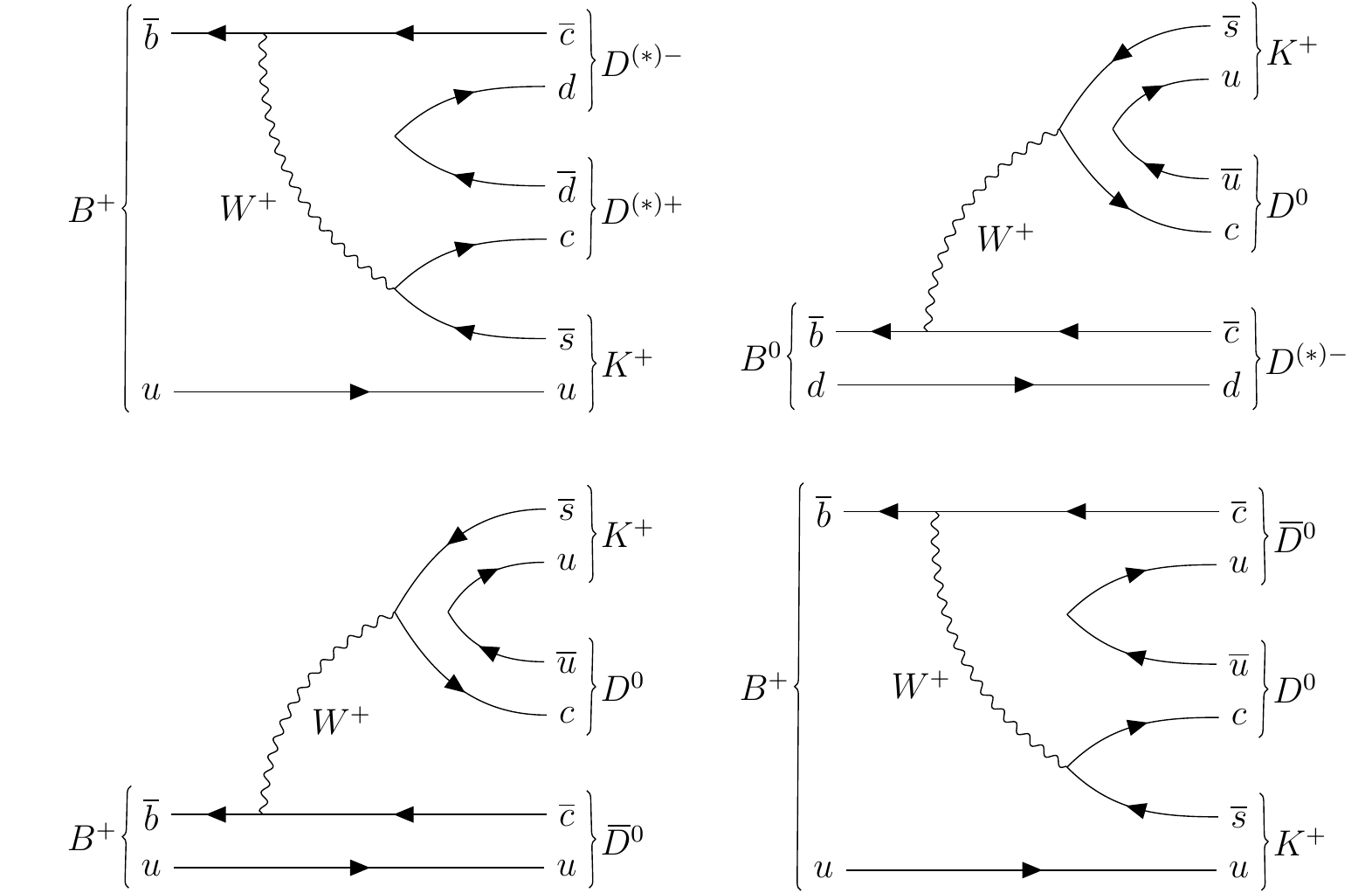}
\caption{Top left: internal \W -emission diagram for the decays \BpDstmDpKp and \BpDstpDmKp. Top right: external \W -emission diagram for the decays \BzDstmDzKp and \BzDmDzKp. Bottom row: (left) external and (right) internal \W -emission diagrams contributing to the \BpDzbarDzKp decay. }
\label{fig:feynman}
\end{figure}

This paper reports measurements of relative branching fractions of \BpDstmDpKp, \BpDstpDmKp, and \BzDstmDzKp decays with respect to the \BpDzbarDzKp decay for the first two, and the \BzDmDzKp decay for the third mode. The decays used for normalisation are chosen due to their similarity to the signal decays in multiplicity and topology, providing the best cancellation of systematic uncertainties on the ratio. Additionally, a relative branching fraction of the \BpDstmDpKp and \BpDstpDmKp decays is reported. The analysis is based on a sample of \proton\proton collisions corresponding to a total integrated luminosity of $9\invfb$ collected at centre-of-mass energies of 7, 8\tev (Run~1), and 13\tev (Run~2) by the \lhcb experiment. The modes containing the excited \Dstar meson are hereafter collectively denoted as \BDstDK and the modes containing only pseudoscalar \D mesons as \BDDK. Decays of these types can proceed at the tree level via three different processes: pure external \W emission, pure internal \W emission, also called colour-suppressed, and the interference of both. Figure~\ref{fig:feynman} shows tree-level diagrams for the processes relevant for this analysis. 

The decays of type $\B \to \Dstarpar \Dstarparb K$ also allow for spectroscopy studies through their intermediate resonant structures, especially for investigations of \cquark\squarkbar resonances via the \Dstarpar\kaon system and charmonium resonances via the \Dstarpar\Dstarparb system. The specific topology of these decays allows for strong suppression of combinatorial background in fully reconstructed decays, and the small energy release leads to an excellent \B-mass resolution. These features make them good candidates for future amplitude analyses. To date, only two amplitude analyses~\cite{Brodzicka:2007aa,Lees:2014abp} have been performed in this family of decays, none of which involved an excited \Dstar meson. Furthermore, both of them are sensitive only to resonant states with natural spin-parity assignments, \ie $J^P = 0^+, 1^-, 2^+, 3^-$, \etc Relatively little is known about states with unnatural spin-parity, and $\B \to \Dstar \Db K$ decays provide an interesting probe for their study.

\section{Detector and simulation}
\label{sec:Detector}

The \lhcb detector~\cite{LHCb-DP-2008-001,LHCb-DP-2014-002} is a single-arm forward spectrometer covering the \mbox{pseudorapidity} range $2<\eta <5$, designed for the study of particles containing \bquark or \cquark quarks. The detector includes a high-precision tracking system consisting of a silicon-strip vertex detector surrounding the $pp$ interaction region, a large-area silicon-strip detector located upstream of a dipole magnet with a bending power of about $4{\mathrm{\,Tm}}$, and three stations of silicon-strip detectors and straw drift tubes placed downstream of the magnet. The tracking system provides a measurement of the momentum, \ptot, of charged particles with a relative uncertainty that varies from 0.5\% at low momentum to 1.0\% at 200\gevc. The minimum distance of a track to a primary \proton\proton collision vertex (PV), the impact parameter (IP),  is measured with a resolution of $(15+29/\pt)\mum$, where \pt is the component of the momentum transverse to the beam, in\,\gevc. Different types of charged hadrons are distinguished using information from two ring-imaging Cherenkov detectors.  Photons, electrons and hadrons are identified by a calorimeter system consisting of scintillating-pad and preshower detectors, an electromagnetic and a hadronic calorimeter. Muons are identified by a system composed of alternating layers of iron and multiwire proportional chambers.

The datasets employed correspond to integrated luminosities of 3\,\invfb and 6\,\invfb, collected during LHC Run 1 (2011 and 2012) and Run 2 (2015--2018). The online event selection is performed by a trigger, which consists of a hardware stage, based on information from the calorimeter and muon systems, followed by a software stage at which the full event is reconstructed. Events passing the hardware trigger are considered in two categories: one in which the trigger criteria are satisfied by energy deposits in the calorimeter associated with the signal candidate decay, and a second in which any of the various muon or calorimeter trigger criteria are met by activity independent of that decay. The software trigger stage requires a two-, three- or four-track secondary vertex with a significant displacement from any primary $pp$ interaction vertex. At least one charged particle must have a transverse momentum $\pt > 1.6\gevc$ and be inconsistent with originating from a PV. A multivariate algorithm~\cite{BBDT,LHCb-PROC-2015-018} is used for the identification of secondary vertices consistent with the decay of a \bquark hadron. 

Simulated samples are produced to model the effect of the detector acceptance and selection requirements, and to guide subsequent fits to the data. To produce these samples, $pp$ collisions are generated using \pythia~\cite{Sjostrand:2007gs,*Sjostrand:2006za}  with a specific \lhcb configuration~\cite{LHCb-PROC-2010-056}. Decays of unstable particles are described by \evtgen~\cite{Lange:2001uf}, in which final-state radiation is generated using \photos~\cite{Golonka:2005pn}. The interaction of the generated particles with the detector, and its response, are implemented using the \geant toolkit~\cite{Allison:2006ve, *Agostinelli:2002hh} as described in Ref.~\cite{LHCb-PROC-2011-006}.

\section{Selection}
\label{sec:Selection}
For this analysis, \Dp mesons are reconstructed via their decay to the $\Km \pip \pip$ final state, and \Dz mesons are reconstructed through their decays to both the $\Km \pip$, denoted as \DzKpi, and $\Km \pip \pip \pim$, denoted as \DzKtpi, final states. However, for decays involving two \Dz mesons at least one must be reconstructed via the two-body decay. The \Dstarp meson is reconstructed through its decay to \Dz\pip, and is labelled as \DstarpKpi (\DstarpKtpi) if decaying into \DzKpi\pip (\DzKtpi\pip). The decays analysed are summarised in Table~\ref{tab:modes}. 
\begin{table}[tpb]
\centering
\caption{Decays under study. In the first column no assumption about the \D final state is made. In the second column, however, the particular \D decays are specified.}
\label{tab:modes}
\begin{tabular}{l | l }
\toprule
  Decay channel & Studied mode  \\
\midrule
\multirow{2}{*} {\BpDstpDmKp}	&	\BpDstpKpiDmKp  \\
								&	\BpDstpK3piDmKp  \\

\midrule

\multirow{2}{*} {\BpDstmDpKp}	&	\BpDstmKpiDpKp  \\
								&	\BpDstmK3piDpKp  \\

\midrule

\multirow{3}{*} {\BzDstmDzKp}	&	\BzDstmKpiDzKpiKp  \\
								&	\BzDstmK3piDzKpiKp  \\
								&	\BzDstmKpiDzK3piKp  \\

\midrule

\multirow{2}{*} {\BpDzbarDzKp}  &	\BpDzbarK3piDzKpiKp  \\
								&	\BpDzbarKpiDzK3piKp  \\

\midrule

\multirow{2}{*} {\BzDmDzKp}		&	\BzDmDzKpiKp  \\
								&	\BzDmDzK3piKp  \\
\bottomrule
\end{tabular}
\end{table}

Well-reconstructed final-state tracks are required. A standard threshold for the \chisqip of each track is applied ($>4$), where \chisqip is defined as the difference in the vertex-fit \chisq for the PV associated with the \B-meson candidate when it is reconstructed with or without the track under consideration. The PV that fits best to the flight direction of the \B candidate is taken as the associated PV. All charged final-state particles must have momentum greater than  $1\gevc$ and transverse momentum above $0.1 \gevc$. At least one of them must have  $p > 10 \gevc$ and $\pt > 1.7 \gevc$, whilst also having an impact parameter with respect to the \B candidate associated PV of at least $0.1\mm$.  The invariant masses of \D candidates are required to lie within $20$\mevcc of their known values~\cite{PDG2019} and their decay vertices must be well reconstructed, having a fit $\chi^2$ less than 10. The \B (\D) candidates have to satisfy the requirement that the minimum of the cosine of the angle between their reconstructed momentum and the line connecting their production and decay vertices should be greater than 0.999 (0). The flight time (distance $\chi^2$) from the associated PV for the \B- (\D)-meson candidates is required to exceed 0.2\,\ps (36). Finally, particle identification (PID) information is employed to aid distinction of final-state \kaon and \pion mesons. The simulated PID response is corrected in order to match the data. This is achieved using calibration $\Dstarp \to \Dz \pip$ samples as a function of track kinematics and multiplicity. An unbinned method is employed, where the probability density functions are modelled using kernel density estimation~\cite{Poluektov:2014rxa}. 

A Boosted Decision Tree (BDT)~\cite{Breiman,AdaBoost} classifier is used to further reduce combinatorial background, consisting of random combinations of tracks that mimic the signal. The BDT is trained using a simulated sample to represent signal and data from the upper sideband of the reconstructed \B-candidate invariant-mass distribution to represent combinatorial background. The variables entering the BDT are: the quality of the reconstructed \B- and \D-meson decay vertices; the \chisqip of the \B- and \D-meson candidates, as well as the \chisqip of the \D-meson decay products; and the particle identification variables of the final-state \kaon and \pion mesons. The threshold for the obtained BDT response is set by optimising the significance of the \B meson signal yield in a fit to data. The signals are sufficiently large that this approach is found to introduce no significant bias to the results. Consistency, within statistical uncertainties, is seen between simulated samples and signal-weighted data for the variables used by the BDT, and the BDT response itself.

A significant peaking background arises from \B-meson decays where the final state is the same but which proceed without one or both of the intermediate charm mesons. The level of this background is estimated by performing a fit to the invariant mass for \B candidates where the reconstructed mass of one or both \D-meson candidates lies far from the known mass and extrapolating the obtained \B signal yield into the \D-meson signal regions. To suppress contributions from these decays, the reconstructed \D-meson decay vertex is required to be downstream of the reconstructed \B-meson decay vertex and a lower bound is placed on the flight distance significance along the beam axis for \D mesons. This requirement suppresses the peaking background to the level of a few percent of the signal yield, and this remaining contamination is later subtracted.

\section{Mass fit}
\label{sec:mass_fit}

\begin{figure}[h!tbp]
\centering
\includegraphics[width=0.44\linewidth]{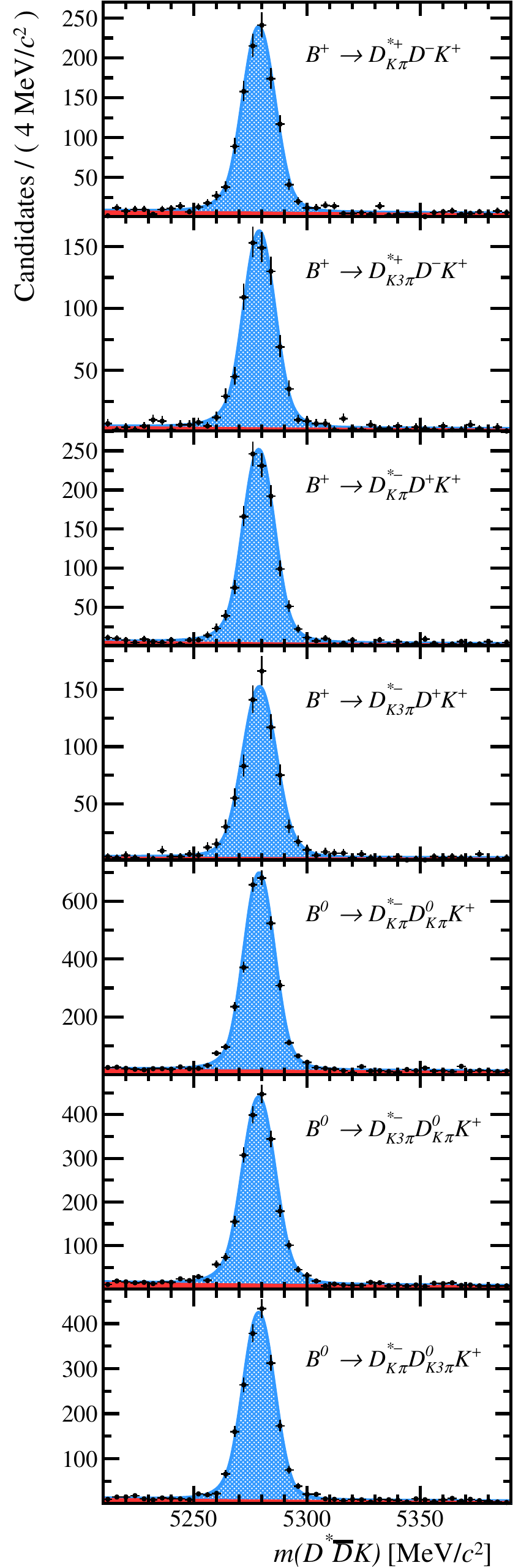}
\includegraphics[width=0.44\linewidth]{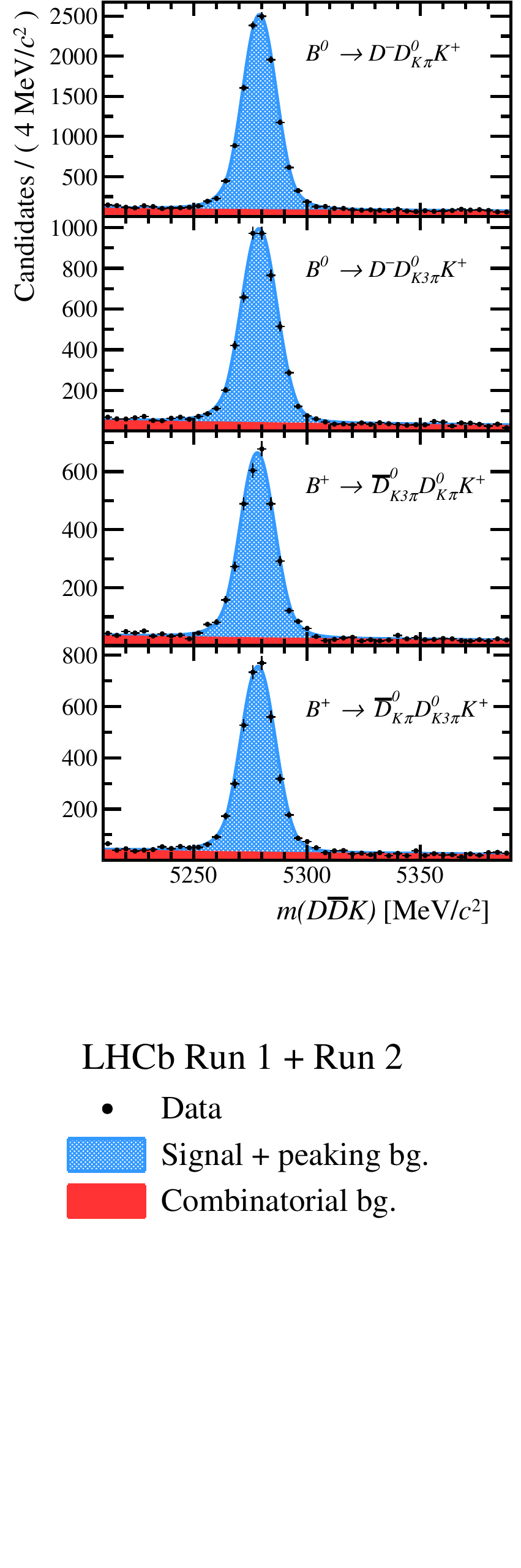}
\caption{Fits to the invariant-mass distributions $m(\Dstarpar\Dbar\kaon)$ of (left) \BDstDK and (right) \BDDK for the combined Run 1 and Run 2 samples. The stacked components are (red) combinatorial background and (blue) signal shape.}
\label{fig:massfit}
\end{figure}

After selecting the signal candidates an unbinned extended maximum-likelihood fit is performed to the distribution of reconstructed \B-candidate mass, $m(\Dstarpar\Dbar\kaon)$, where the reconstruction is performed with \D-candidate masses constrained to their known values~\cite{PDG2019} and the \B-candidate direction of flight to be originating at the PV. The fit to the mass distribution is performed in the range from $5210$ to $5390 \mevcc$, separately for Run 1 and Run 2 data. The shape used to fit the distribution consists of two components: one to describe the decays of a signal $B$ meson, and a second to model the combinatorial background. The signal shape is modelled using a Double-Sided Crystal Ball (DSCB)~\cite{Skwarnicki:1986xj} function. The asymmetric shape and non-Gaussian tails account for both the mass-resolution effects on both sides and energy loss due to final-state radiation. The values of tail parameters of the DSCB shapes are fixed to those found in simulated decays while the Gaussian core parameters are extracted from the fit together with the signal yield. To model the combinatorial background an exponential function is used. The lower bound on the range of invariant mass considered excludes any significant background from partially reconstructed decays. The combined Run 1 and Run 2 invariant-mass distributions and fit results are shown in Fig.~\ref{fig:massfit}. The fit is used to extract a signal weight for each candidate using the \sPlot technique~\cite{Pivk:2004ty}. 

\section{Efficiencies}
\label{sec:Efficiencies}

The efficiencies $\varepsilon$ of the selection of signal candidates are calculated separately for Run 1 and Run 2 in two stages: 
\begin{equation}
    \varepsilon = \varepsilon^{\mathrm{acc}} \cdot \varepsilon^{\mathrm{sel}},
\end{equation}
where the geometric \lhcb acceptance efficiencies $\varepsilon^{\mathrm{acc}}$ are calculated using simulated samples, and correspond to the fraction of generated events where all final-state particles lie within the \lhcb acceptance. The trigger, reconstruction, and selection efficiencies $\varepsilon^{\mathrm{sel}}$ are also determined using simulated samples as the fraction of reconstructed candidates passing the trigger, reconstruction, and selection criteria, given that they pass the geometrical acceptance requirement. The efficiencies are evaluated as a function of the position in the phase space of the decay. Due to the presence of a pseudoscalar particle in the initial state and one vector (\Dstar) plus two pseudoscalar particles in the final state, decays of the type \BDstDK have four independent degrees of freedom. These are chosen to be the two-body squared invariant masses $m^2(\Dstar\kaon)$ and $m^2(\Dbar\kaon)$, and two helicity angles: the angle $\chi$ between the decay planes of the \Dstar meson and the $\Dbar \kaon$ system in the \B-meson rest frame, and the \Dstar-meson helicity angle $\theta$ defined as the angle between the direction of the \pion meson coming from the \Dstar meson in the \Dstar-meson rest frame, and the \Dstar meson in and \B-meson rest frame. In the case of \BDDK decays only two degrees of freedom are required, and these are chosen to be the two-body squared invariant masses $m^2(\D\kaon)$ and $m^2(\Dbar\kaon)$. 

\begin{figure}[h!tbp]
\centering
\includegraphics[width=0.91\linewidth]{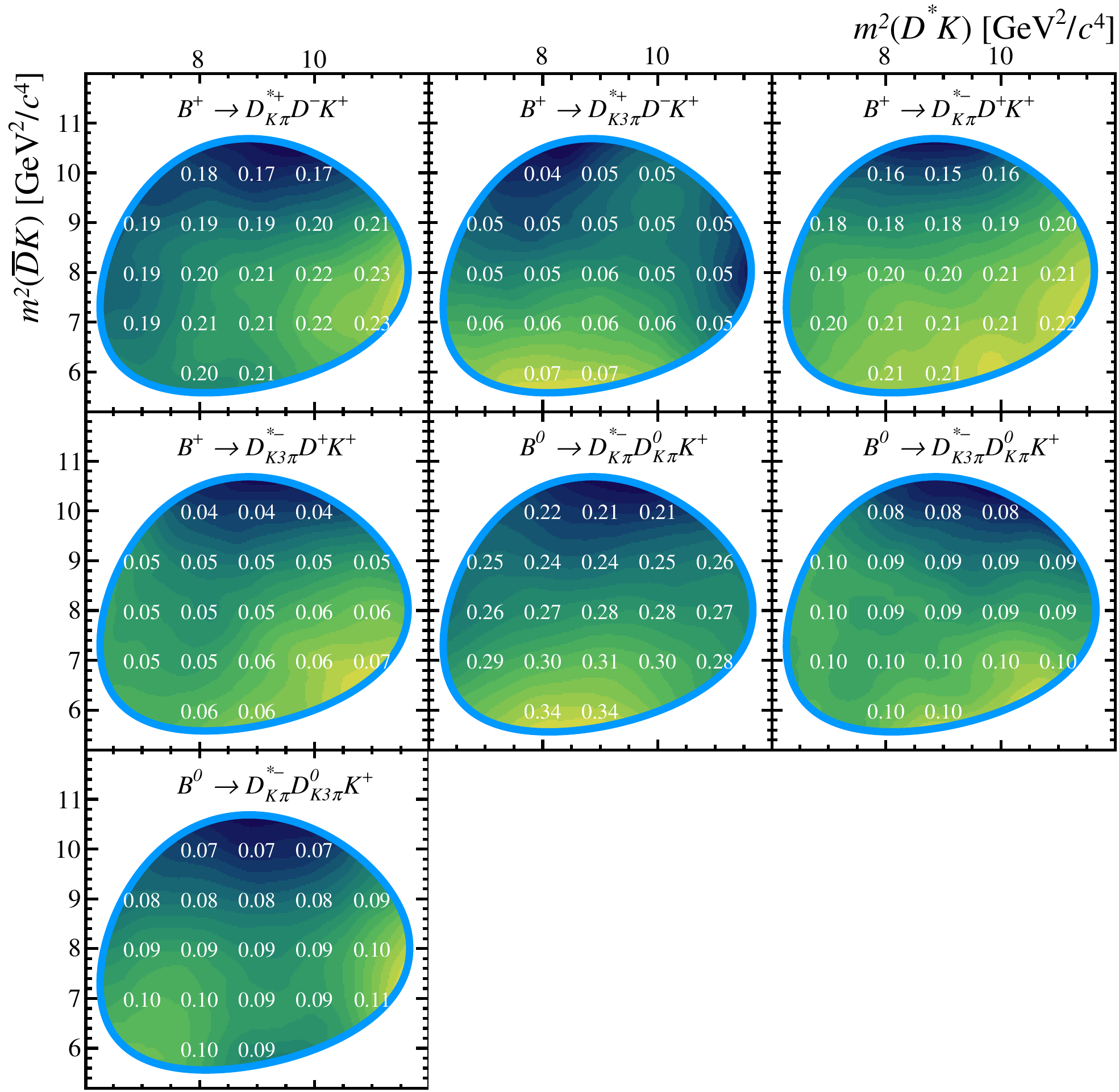}
\llap{\shortstack{%
        \includegraphics[width=0.5\linewidth]{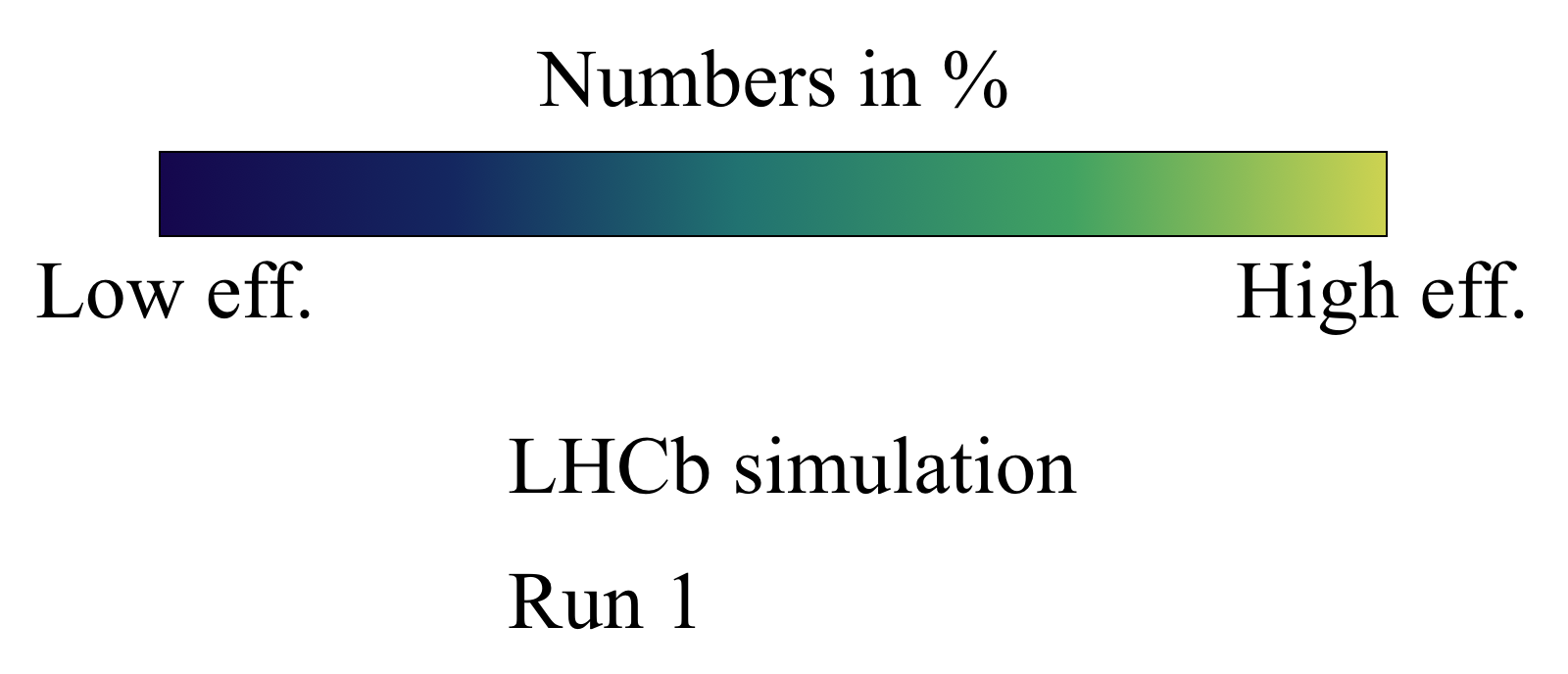}\\
        \rule{0ex}{0.25in}%
      }
  \rule{0.25in}{0ex}}
\caption{Selection and reconstruction efficiency, $\varepsilon^{\mathrm{sel}}$, as a function of position in the two-body squared invariant-mass plane for the seven \BDstDK modes, obtained using Run~1 simulated samples. A KDE smoothing has been applied. The blue lines indicate the kinematic boundaries and the numbers indicate the value of the efficiency at several points in the phase space.}
\label{fig:efficiency_run1}
\end{figure}

\begin{figure}[h!tbp]
\centering
\includegraphics[width=0.91\linewidth]{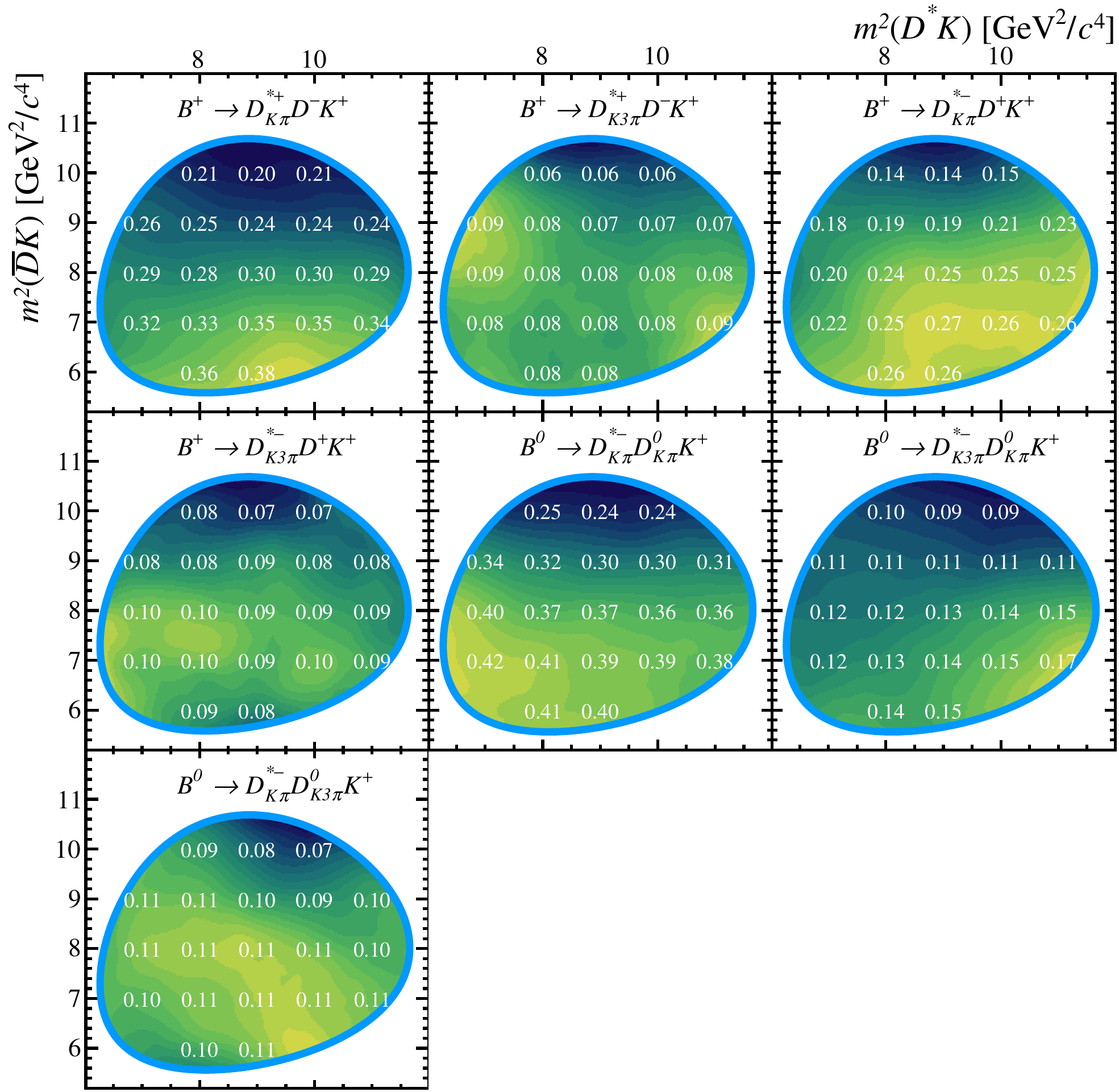}
\llap{\shortstack{%
        \includegraphics[width=0.5\linewidth]{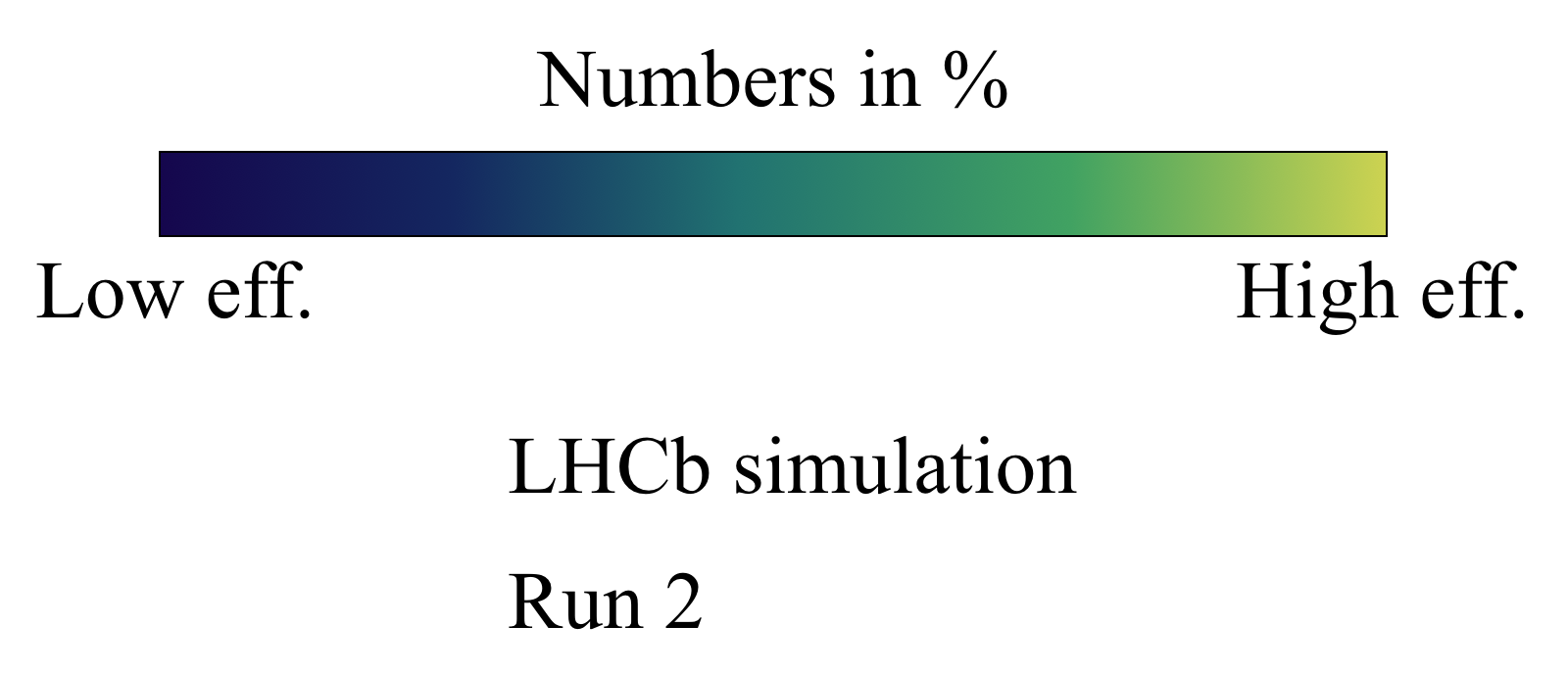}\\
        \rule{0ex}{0.25in}%
      }
  \rule{0.25in}{0ex}}
\caption{Selection and reconstruction efficiency, $\varepsilon^{\mathrm{sel}}$, as a function of position in the two-body squared invariant-mass plane for the seven \BDstDK modes, obtained using Run~2 simulated samples. A KDE smoothing has been applied. The blue lines indicate the kinematic boundaries and the numbers indicate the values of the efficiency at several points in the phase space.}
\label{fig:efficiency_run2}
\end{figure}

Whilst the efficiency varies considerably across the two-body invariant-mass planes and the \Dstar-meson helicity angle $\theta$, it does not depend significantly on the angle $\chi$. Two-dimensional efficiency distributions, as functions of $m^2(\Dstar\kaon)$ and $m^2(\Dbar\kaon)$, are obtained in four equal bins of $\cos(\theta)$. The efficiency distributions are further smoothed using a kernel density estimation (KDE) technique~\cite{Poluektov:2014rxa}. The efficiency in the two-body invariant-mass distribution integrated over the two helicity angles are shown in Figs.~\ref{fig:efficiency_run1} and~\ref{fig:efficiency_run2} for the \BDstDK samples from Run 1 and Run 2, respectively. The relative statistical uncertainties on the total efficiencies are in range $10-20\%$.

\section{Corrected yields}
\label{sec:yields}

The ratios of branching fractions are calculated using signal yields corrected by applying candidate-by-candidate background subtraction and efficiency correction, and accounting for the decays of the \D mesons into the final states. The branching fraction of a $\B \to \Dstarpar \Db K$ decay is proportional to the corrected yield, \Ncorr, calculated as
\begin{equation}
\label{eq:corr_yield}
\Ncorr = \frac{\displaystyle \sum_{i} \frac{W_i}{\epsilon^{\mathrm{sel}}_i (x_i) \cdot \epsilon^{\mathrm{acc}}} - n^{\mathrm{corr}}_{\mathrm{peaking}}}{\displaystyle \BF(\Dstarpar)\cdot\BF(\Dbar)}.
\end{equation}
Here the index $i$ runs over all candidates in the fitted sample, $W_i$ is the signal weight for candidate $i$ (see Section~\ref{sec:mass_fit}), $\epsilon_i^{\mathrm{sel}}$ is the selection efficiency for candidate $i$ as a function of its position $x_i$ in the relevant phase space, and $\epsilon^{\mathrm{acc}}$ is the efficiency of the acceptance cut for the given mode (see Section~\ref{sec:Efficiencies}). Since the efficiency-weighted sum over candidates includes a small (peaking) background contribution, the efficiency-corrected residual peaking background $n^{\mathrm{corr}}_{\mathrm{peaking}}$ is subtracted from the signal region. The value of $n^{\mathrm{corr}}_{\mathrm{peaking}}$ is obtained by taking the estimated yield of the peaking background and dividing it by an average efficiency of the sample, since the distribution of the peaking background in the phase space of the decay is not known. Finally, the denominator is used to correct for the \D-meson decay branching fractions, which are:
\begin{equation*}
\begin{alignedat}{3}
\BF &(\Dz \rightarrow \Km \pip) &&= (3.999\pm0.045)\%\quad &&\text{~\cite{Amhis:2019ckw}}, \\
\BF &(\Dz \rightarrow \Km \pip \pip \pim) &&= (8.23\pm0.14)\%\quad &&\text{~\cite{PDG2019}}, \\
\BF &(\Dp \rightarrow \Km \pip \pip) &&= (9.38\pm0.16)\%\quad &&\text{~\cite{PDG2019}}, \\
\BF &(\Dstarp \rightarrow \Dz \pip) &&= (67.7\pm0.5)\%\quad &&\text{~\cite{PDG2019}}. 
\end{alignedat}
\end{equation*}

Table~\ref{tab:yields_all} summarises the values of signal yields $N$ obtained from the mass fits as well as the corrected yields \Ncorr for all studied modes.

\renewcommand{\arraystretch}{1.15}

\begin{table}[tbp]
\centering
\caption{Table of all signal yields $N$ and efficiency and \D-meson branching fraction corrected yields \Ncorr with the residual peaking background subtracted. The values of corrected yields are rounded to the order of $10^6$. The uncertainties are statistical only.}
\label{tab:yields_all}
\begin{tabular}{@{}l|r@{$\,\pm\,$}lr@{$\,\pm\,$}l|r@{$\,\pm\,$}lr@{$\,\pm\,$}l@{}}
\toprule
\multirow{2}{*}{Mode} & \multicolumn{4}{c|}{Run 1} & \multicolumn{4}{c}{Run 2} \\
                       & \multicolumn{2}{c}{$N$}    &  \multicolumn{2}{c|}{$\Ncorr\,(10^6)$} & \multicolumn{2}{c}{$N$}     &  \multicolumn{2}{c}{$\Ncorr\,(10^6)$}   \\ \midrule
\BpDstpKpiDmKp & $212$ & $16$ &  $289$ & $21$ & $869$ & \phantom{0}$32$  & $854$ & \phantom{0}$32$ \\
\BpDstpK3piDmKp & $116$ & $11$  & $286$ & $28$ & $606$ & \phantom{0}$26$  & $997$ & \phantom{0}$44$ \\
\BpDstmKpiDpKp & $210$ & $15$ &  $313$ & $23$ & $912$ & \phantom{0}$32$  & $1009$ & \phantom{0}$36$ \\
\BpDstmK3piDpKp & $153$ & $13$ &  $371$ & $32$ & $566$ & \phantom{0}$25$ & $969$ & \phantom{0}$45$ \\
\BzDstmKpiDzKpiKp & $605$ & $26$ &  $1196$ & $52$ & $2409$ & \phantom{0}$52$  & $3495$ & \phantom{0}$76$ \\
\BzDstmK3piDzKpiKp & $321$ & $20$ &  $949$ & $57$ & $1706$ & \phantom{0}$44$  & $3541$ & \phantom{0}$92$ \\
\BzDstmKpiDzK3piKp & $331$ & $20$ &  $1105$ & $64$ & $1544$ & \phantom{0}$41$ & $3812$ & $104$ \\
\BpDzbarK3piDzKpiKp & $477$ & $24$ &  $517$ & $26$ & $2564$ & \phantom{0}$56$ & $1823$ & \phantom{0}$39$ \\
\BpDzbarKpiDzK3piKp & $622$ & $28$ &  $527$ & $23$ & $2853$ & \phantom{0}$60$ & $1720$ & \phantom{0}$35$ \\
\BzDmDzKpiKp & $2443$ & $54$ &  $651$ & $14$ & $9071$ & $104$ & $2039$ & \phantom{0}$23$ \\
\BzDmDzK3piKp & $864$ & $32$ &  $648$ & $23$ & $3867$ & \phantom{0}$69$ & $2040$ & \phantom{0}$36$ \\
\bottomrule
\end{tabular}
\end{table}

\section{Systematic uncertainties}
\label{sec:syst}

Many systematic effects cancel exactly in the ratios of branching fractions, such as the uncertainties in the \bbbar-production cross-section and fragmentation fractions as well as the uncertainties in the luminosity. The kinematics differ most between numerator and denominator for the slow pion in modes involving a \Dstar decay, but the tracking efficiency of the slow pion produced in the \Dstar decay is found to be well modelled using calibration samples and the associated systematic uncertainty is found to be negligible.  Uncertainties are considered where they arise from the shapes used to model the invariant-mass distribution, the efficiency determination, the resampling of the PID response, and the contribution of residual peaking backgrounds.

The systematic uncertainty related to the signal model is evaluated by randomly sampling each tail parameter of the DSCB from a normal distribution centred at the value used in the fit and with a width corresponding to its uncertainty. The fit is then repeated with these new values and the yields are recalculated. The correlations of the tail parameters are accounted for. By doing this many times a distribution of yields is obtained. The RMS of this distribution is then used as the systematic uncertainty. Changing the shape of the background model is found to have a negligible impact on the resulting yields. The associated systematic uncertainty is thus neglected.

To estimate the systematic uncertainty associated with the choice of the kernel width in the PID response correction, the procedure is repeated with a larger kernel width. The absolute difference between the new efficiency-corrected yield and the baseline value is taken as the uncertainty. 

Even after applying the flight-distance significance requirements on the \D mesons there is still some underlying residual peaking background $n^{\mathrm{corr}}_{\mathrm{peaking}}$. This is subtracted from the signal yield. The uncertainty on the yield of the residual peaking background, determined using the $c$-hadron sidebands, is used as the systematic uncertainty.

The limited size of the simulated samples leads to uncertainties in the efficiency estimations. Bootstrapped samples are produced by sampling randomly candidates from the original simulated sample, allowing repeated selection of the same candidate, until a new sample having the same number of candidates is derived. These samples are used to evaluate the associated systematic uncertainty, resulting in an ensemble of different efficiency distributions. The RMS values of the resulting yield distributions are then taken as a measure of the systematic uncertainties. This is typically the dominant systematic uncertainty.

The tracking efficiencies are assumed to cancel in all ratios where the same number of tracks is reconstructed in the numerator and denominator. Differences in kinematics, most obviously for the slow pion in the \Dstar decay, could lead to imperfect cancellation. This was explored and the effect was found to be negligible. In ratios where the number of tracks differ in the numerator and denominator, an additional systematic uncertainty of 1\% per additional track is applied.

The magnitudes of the individual contributions are summarised in Table~\ref{tab:syst} together with the total systematic uncertainty obtained by combining the individual components in quadrature.

\renewcommand{\arraystretch}{1.18}

\begin{table}[tbp]
\centering
\caption{Systematic uncertainties on \Ncorr from the signal PDF parameters ($\sigma_{\mathrm{PDF}}$), the finite simulation samples ($\sigma_{\mathrm{MC}}$), the PID resampling ($\sigma_{\mathrm{PID}}$), the residual peaking background ($\sigma_{\mathrm{bkg}}$), and the total systematic uncertainty ($\sigma_{\mathrm{tot.}}$). All values are given as a percentage of the central value of \Ncorr.}
\label{tab:syst}
\begin{tabular}{@{}l | cccc|c | cccc | c@{}}
\toprule
\multirow{2}{*}{Decay channel} & \multicolumn{5}{c|}{Run 1 (\%)} & \multicolumn{5}{c}{Run 2 (\%)} \\
 & $\sigma_{\mathrm{PDF}}$ & $\sigma_{\mathrm{MC}}$ & $\sigma_{\mathrm{PID}}$ & $\sigma_{\mathrm{bkg}}$ & $\sigma_{\mathrm{tot.}}$ & $\sigma_{\mathrm{PDF}}$ & $\sigma_{\mathrm{MC}}$ & $\sigma_{\mathrm{PID}}$ & $\sigma_{\mathrm{bkg}}$ & $\sigma_{\mathrm{tot.}}$ \\
\midrule
\BpDstpKpiDmKp & $0.6$ & $0.8$ & $1.5$ & $0.8$ & $2.0$ & $0.5$ & $1.4$ & $0.2$ & $0.5$ & $1.6$ \\
\BpDstpK3piDmKp & $1.2$ & $1.2$ & $0.9$ & $1.4$ & $2.4$ & $1.0$ & $2.1$ & $0.7$ & $0.6$ & $2.5$ \\
\BpDstmKpiDpKp & $0.5$ & $1.0$ & $0.4$ & $0.7$ & $1.4$ & $0.8$ & $1.8$ & $0.7$ & $0.4$ & $2.1$ \\
\BpDstmK3piDpKp & $1.4$ & $1.6$ & $1.1$ & $1.2$ & $2.7$ & $0.7$ & $2.5$ & $1.2$ & $0.6$ & $2.9$ \\
\BzDstmKpiDzKpiKp & $0.6$ & $0.7$ & $0.9$ & $0.3$ & $1.3$ & $0.5$ & $1.1$ & $0.2$ & $0.2$ & $1.2$ \\
\BzDstmK3piDzKpiKp & $0.8$ & $1.2$ & $0.3$ & $0.7$ & $1.6$ & $0.8$ & $1.7$ & $0.6$ & $0.3$ & $2.0$ \\
\BzDstmKpiDzK3piKp & $0.9$ & $1.2$ & $0.3$ & $0.6$ & $1.6$ & $0.6$ & $2.0$ & $0.3$ & $0.3$ & $2.1$ \\
\BpDzbarK3piDzKpiKp & $0.6$ & $1.1$ & $1.0$ & $0.9$ & $1.8$ & $1.1$ & $1.8$ & $0.5$ & $0.4$ & $2.2$ \\
\BpDzbarKpiDzK3piKp & $0.7$ & $1.1$ & $0.5$ & $0.7$ & $1.6$ & $0.7$ & $1.6$ & $0.4$ & $0.3$ & $1.8$ \\
\BzDmDzKpiKp & $0.4$ & $0.7$ & $0.5$ & $0.4$ & $1.0$ & $0.3$ & $0.7$ & $0.7$ & $0.2$ & $1.1$ \\
\BzDmDzK3piKp & $0.2$ & $1.4$ & $0.3$ & $0.5$ & $1.5$ & $0.8$ & $1.3$ & $0.4$ & $0.3$ & $1.6$ \\
\bottomrule
\end{tabular}
\end{table}

\section{Results}
\label{sec:Results}

The ratios of branching fractions are obtained by appropriately combining the \Ncorr yields of decay modes in Table~\ref{tab:modes} into ratios, such that the systematic uncertainty coming from the different number of tracks in the numerator and denominator is minimised. 
In order to calculate the first two branching-fraction ratios of the \BpDstmDpKp(\BpDstpDmKp) decay with respect to the \BpDzbarDzKp decay a weighted average of \Ncorr of \BpDstmKpiDpKp(\BpDstpKpiDmKp) and \BpDstmKtpiDpKp(\BpDstpKtpiDmKp) is done and divided by the weighted average of \Ncorr for the \BpDzbarKtpiDzKpiKp and \BpDzbarKpiDzKtpiKp modes. The associated weight in the weighted average is the inverse of the variance of the value. The variance on \Ncorr is obtained by adding the statistical and the systematic uncertainty, including the uncertainties due to \D-meson branching fractions, in quadrature. 

The first measurement of the third ratio of \BzDstmDzKp to  \BzDmDzKp decays is calculated by performing a weighted average of \Ncorr for \BzDstmKtpiDzKpiKp and \BzDstmKpiDzKtpiKp decays, and dividing it by the value of \Ncorr for the \BzDmDzKtpiKp decay. A second measurement is obtained by finding the ratio of \Ncorr for \BzDstmKpiDzKpiKp and \BzDmDzKpiKp, which is combined with the first one into the final branching-fraction ratio.

The fourth branching-fraction ratio of \BpDstmDpKp and \BpDstpDmKp decays is calculated as the weighted average of two ratios. The first is the ratio of \BpDstpKpiDmKp and \BpDstmKpiDpKp decays, and the second is that for \BpDstpKtpiDmKp and \BpDstmKtpiDpKp decays.

The ratios of branching fractions are computed separately for Run~1 and Run~2 and then combined in a weighted average. These ratios are measured to be 
\begin{equation*}
\begin{split}
\frac{\BF (\BpDstpDmKp)}{\BF (\BpDzbarDzKp)} &= 0.517 \pm 0.015 \pm 0.013 \pm 0.011  , \\[10pt]
\frac{\BF (\BpDstmDpKp)}{\BF (\BpDzbarDzKp)} &= 0.577 \pm 0.016 \pm 0.013 \pm 0.013  , \\[10pt]
\frac{\BF (\BzDstmDzKp)}{\BF (\BzDmDzKp)}    &= 1.754 \pm 0.028 \pm 0.016 \pm 0.035  , \\[10pt]
\frac{\BF (\BpDstpDmKp)}{\BF (\BpDstmDpKp)}  &= 0.907 \pm 0.033 \pm 0.014  ,
\end{split}
\end{equation*}
where the first uncertainty is statistical, the second systematic, and the third one is due to the uncertainties on the \D-meson branching fractions~\cite{PDG2019}. 

The BaBar collaboration studied these decays previously~\cite{delAmoSanchez:2010pg}, with a different set of $D^{*0}$ and $\Dz$ channels, obtaining signal yields of $91\pm13$ \BpDstpDmKp candidates, $75\pm13$ \BpDstmDpKp candidates, and $1300\pm54$ \BzDstmDzKp candidates. The sizes of the signal yields obtained using the LHCb data are around twenty times larger for the first two decays, and over five times larger for the third. Significant increases are seen for the yields obtained in the normalisation modes, with respect to earlier studies using data from the Belle and BaBar experiments. Good agreement is seen with respect to the corresponding branching fraction ratios according to the Particle Data Group (PDG)~\cite{PDG2019}, calculated to be $0.43\pm0.12$, $0.41\pm0.13$, $2.3\pm0.3$, and $1.1\pm0.3$, respectively. The measurements described in this article are between 5 and 7 times more precise. The ratio between the \BpDstpDmKp and \BpDstmDpKp deviates from unity with a significance just below $3\sigma$, suggesting activity in a channel other than the $\Dp^{*}\Dm$ channel that the two have in common. These measurements, and the high purity of the samples obtained for the decays under study, make these decays prime targets for future analyses of resonant structure.

\section{Summary}
\label{sec:summary}

A data sample corresponding to an integrated luminosity of $9\invfb$ recorded with the \lhcb detector is used to measure four ratios of branching fractions in $\B \to \Dstarpar \Db K$ decays. The ratios are consistent with previous measurements and are measured with the highest precision to date. Furthermore, this work represents the first published analysis at the \lhc of $b$-hadron decays to two open-charm hadrons and a third, light, hadron. Large samples of $\B \to \Dstarpar \Db K$ decays are available, and can be isolated in the \lhcb dataset with low background contamination. These are promising characteristics for these channels with future studies of their intermediate resonant structure in view.

\section*{Acknowledgements}
%
%
\noindent We express our gratitude to our colleagues in the CERN
accelerator departments for the excellent performance of the LHC. We
thank the technical and administrative staff at the LHCb
institutes.
We acknowledge support from CERN and from the national agencies:
CAPES, CNPq, FAPERJ and FINEP (Brazil); 
MOST and NSFC (China); 
CNRS/IN2P3 (France); 
BMBF, DFG and MPG (Germany); 
INFN (Italy); 
NWO (Netherlands); 
MNiSW and NCN (Poland); 
MEN/IFA (Romania); 
MSHE (Russia); 
MICINN (Spain); 
SNSF and SER (Switzerland); 
NASU (Ukraine); 
STFC (United Kingdom); 
DOE NP and NSF (USA).
We acknowledge the computing resources that are provided by CERN, IN2P3
(France), KIT and DESY (Germany), INFN (Italy), SURF (Netherlands),
PIC (Spain), GridPP (United Kingdom), RRCKI and Yandex
LLC (Russia), CSCS (Switzerland), IFIN-HH (Romania), CBPF (Brazil),
PL-GRID (Poland) and OSC (USA).
We are indebted to the communities behind the multiple open-source
software packages on which we depend.
Individual groups or members have received support from
AvH Foundation (Germany);
EPLANET, Marie Sk\l{}odowska-Curie Actions and ERC (European Union);
A*MIDEX, ANR, Labex P2IO and OCEVU, and R\'{e}gion Auvergne-Rh\^{o}ne-Alpes (France);
Key Research Program of Frontier Sciences of CAS, CAS PIFI,
Thousand Talents Program, and Sci. \& Tech. Program of Guangzhou (China);
RFBR, RSF and Yandex LLC (Russia);
GVA, XuntaGal and GENCAT (Spain);
the Royal Society
and the Leverhulme Trust (United Kingdom).

\addcontentsline{toc}{section}{References}
\setboolean{inbibliography}{true}
\bibliographystyle{LHCb}
\bibliography{standard,LHCb-PAPER,LHCb-CONF,LHCb-DP,LHCb-TDR,main}
 
\newpage
\centerline
{\large\bf LHCb collaboration}
\begin
{flushleft}
\small
R.~Aaij$^{31}$,
C.~Abell{\'a}n~Beteta$^{49}$,
T.~Ackernley$^{59}$,
B.~Adeva$^{45}$,
M.~Adinolfi$^{53}$,
H.~Afsharnia$^{9}$,
C.A.~Aidala$^{82}$,
S.~Aiola$^{25}$,
Z.~Ajaltouni$^{9}$,
S.~Akar$^{64}$
% ...

{\footnotesize \it

% ...
$ ^{9}$Universit{\'e} Clermont Auvergne, CNRS/IN2P3, LPC, Clermont-Ferrand, France\\
% ...
$ ^{25}$INFN Sezione di Milano, Milano, Italy\\
% ...
$ ^{31}$Nikhef National Institute for Subatomic Physics, Amsterdam, Netherlands\\
% ...
$ ^{45}$Instituto Galego de F{\'\i}sica de Altas Enerx{\'\i}as (IGFAE), Universidade de Santiago de Compostela, Santiago de Compostela, Spain\\
% ...
$ ^{49}$Physik-Institut, Universit{\"a}t Z{\"u}rich, Z{\"u}rich, Switzerland\\
% ...
$ ^{53}$H.H. Wills Physics Laboratory, University of Bristol, Bristol, United Kingdom\\
% ...
$ ^{59}$Oliver Lodge Laboratory, University of Liverpool, Liverpool, United Kingdom\\
% ...
$ ^{64}$University of Cincinnati, Cincinnati, OH, United States\\
% ...
$ ^{67}$Syracuse University, Syracuse, NY, United States\\
% ...
$ ^{82}$University of Michigan, Ann Arbor, United States, associated to $^{67}$\\
\bigskip
% ...
\medskip
}
\end{flushleft}

\end{document}